\documentclass[twocolumn,english,aps,prl,longbibliography,showpacs,amsfonts,amsmath,amssymb,superscriptaddress,
footinbib,floatfix]{revtex4-1}
\usepackage[T1]{fontenc}
\usepackage[latin9]{inputenc}
\usepackage{amstext}
\usepackage{graphicx}

\makeatletter
\usepackage{color}
\usepackage{enumerate}
\def\kbar{{\mathchar'26\mkern-9mu k}}
\newcommand{\rmd}{\mathrm{d}}
\newcommand{\vecq}{\mathbf{q}}

\makeatother

\usepackage{babel}
\begin{document}

\title{Experimental observation of two-dimensional Anderson localization
with the atomic kicked rotor}

\author{Isam Manai}

\affiliation{Univ. Lille, CNRS, UMR 8523 -- PhLAM -- Laboratoire de Physique des Lasers Atomes et Molécules, F-59000 Lille, France}

\homepage{www.phlam.univ-lille1.fr/atfr/cq}

\author{Jean-Fran\c cois Cl{\'e}ment}

\affiliation{Univ. Lille, CNRS, UMR 8523 -- PhLAM -- Laboratoire de Physique des Lasers Atomes et Molécules, F-59000 Lille, France}

\author{Radu Chicireanu}

\affiliation{Univ. Lille, CNRS, UMR 8523 -- PhLAM -- Laboratoire de Physique des Lasers Atomes et Molécules, F-59000 Lille, France}

\author{Cl{\'e}ment Hainaut}

\affiliation{Univ. Lille, CNRS, UMR 8523 -- PhLAM -- Laboratoire de Physique des Lasers Atomes et Molécules, F-59000 Lille, France}

\author{Jean Claude Garreau}

\affiliation{Univ. Lille, CNRS, UMR 8523 -- PhLAM -- Laboratoire de Physique des Lasers Atomes et Molécules, F-59000 Lille, France}

\author{Pascal Szriftgiser}

\affiliation{Univ. Lille, CNRS, UMR 8523 -- PhLAM -- Laboratoire de Physique des Lasers Atomes et Molécules, F-59000 Lille, France}

\author{Dominique Delande}

\affiliation{Laboratoire Kastler Brossel, UPMC, CNRS, ENS, Coll{\`e}ge de France;
4 Place Jussieu, F-75005 Paris, France}

%\date{version 12, November 8, 2015}
\begin{abstract}
Dimension 2 is expected to be the lower critical dimension for Anderson
localization in a time-reversal-invariant disordered quantum system. 
Using an atomic quasiperiodic kicked rotor -- equivalent to a two-dimensional
Anderson-like model -- we experimentally study Anderson localization
in dimension 2 and we observe localized wavefunction dynamics. We
also show that the localization length depends exponentially on the
disorder strength and anisotropy and is in quantitative
agreement with the predictions of the self-consistent theory for the
2D Anderson localization.
\end{abstract}

\pacs{03.75.-b , 72.15.Rn, 05.45.Mt, 64.70.qj}

\maketitle

The metal-insulator Anderson transition plays a central role in the
study of quantum disordered systems. Using a tight-binding description
of an electron in a lattice, Anderson~\citep{Anderson:LocAnderson:PR58}
postulated in 1958 that the dominant effect of impurities in a crystal
is to randomize the diagonal term of the Hamiltonian, and showed that
this may lead to a localization of the wavefunction, in sharp
contrast with the Bloch-wave solution for a perfect crystal. In a
weakly disordered (3D) crystal, the eigenstates are delocalized, leading
to a diffusive (metallic) transport, while strong disorder produces
an insulator with localized eigenstates. From its original solid-state
physics scope \citep{Anderson:LocAnderson:PR58,Altshuler:MetalInsulator:ANP06,Kramer:Localization:RPP93,Thouless:AndLoc:PREP74}
this approach has been applied to a large class of systems in which
waves propagate in disorder. This includes quantum-chaotic systems
\citep{Casati:LocDynFirst:LNP79,Fishman:LocDynAnderson:PRA84} and
electromagnetic radiation \citep{Maret:AndersonTransLight:PRL06,Schwartz:LocAnderson2DLight:N07,Dembowski:AndersonMicrocavity:PRE99}.
Important theoretical progress was obtained in ref.~\cite{Abrahams:Scaling:PRL79},
which postulated that Anderson localization can be described by a
one-parameter scaling law, leading to the prediction that, for $d\le2$,
the dynamics is generically localized, even if the disorder is very
weak. For $d>2$, it predicted the existence of the Anderson
transition between a diffusive dynamics at weak disorder and a localized
dynamics at strong disorder. 

There is no fully quantitative theory of Anderson localization, and analytic results
are scarce. Supersymmetry techniques~\citep{Evers:AndersonTransitions:RMP08}
allow derivation of expansions in powers of $d-2$ of the various
quantities of interest, but reaching even $d=3$ is difficult. A useful, simplified theoretical approach
is the so-called self-consistent theory of
localization. In few words, it can be thought as a mean field theory
where large fluctuations are neglected, but where weak localization
corrections to transport, due to interference between time reversed
multiply scattered paths, are included self-consistently. For spinless
time-invariant systems, belonging to the orthogonal symmetry class~\citep{Evers:AndersonTransitions:RMP08},
this approach correctly predicts the existence of the metal-insulator
Anderson transition for $d>2$, 
although it fails to predict the correct critical exponent. For $d=1$,
it quantitatively predicts the localization length in a weak disorder.
Other approaches lead to approximate values for the critical exponent
not far from the numerical prediction~\cite{Garcia-Garcia:SemiclassicalTheoryAndersonTrans:PRL08}.

Dimension $d=2$ -- the lower critical dimension -- is very special,
the localization properties depending on the symmetry class. In the
orthogonal symmetry class,
the dynamics is always localized, but the localization length is predicted
to scale \emph{exponentially} with the inverse of the disorder strength,
i.e. $\xi\propto\ell\exp(\pi k\ell/2)$ \cite{Kuhn:CohMatterWaveTransportSpeckle:NJP07}
where $k$ is the wavevector and $\ell$ the mean-free path
for propagation in the disordered medium. As discussed in the 
Supplemental Material~\footnote{See Supplemental Material at [URL will be inserted by publisher] for details on the scaling theory and on the self-consistent theory of 2D localization in the kicked rotor.},
such an exponential dependence is a signature of the fact that $d=2$ is the lower critical dimension
for Anderson localization. 
The 2D case has been previously
studied experimentally in optical and ultracold atom systems
~\citep{Schwartz:LocAnderson2DLight:N07,Saint-Vincent:Diffusion2DDisorderUcoldAt:PRL10}, but no quantitative indication of the exponential
scaling has been demonstrated yet. In the present Letter, we use
the well-known correspondence between the $d$-dimension Anderson
model and the $d$-frequency quasiperiodic kicked rotor~\citep{Fishman:LocDynAnderson:PRA84,Shepelyansky:Bicolor:PD87,Casati:IncommFreqsQKR:PRL89}
to test experimentally these predictions.

The quasiperiodic kicked rotor (QPKR)~\citep{Shepelyansky:Bicolor:PD87,Casati:LocDynFirst:LNP79,Fishman:LocDynAnderson:PRA84,Moore:AtomOpticsRealizationQKR:PRL95,Casati:IncommFreqsQKR:PRL89}
is a spatially one-dimensional system with an engineered time-dependence
such that its dynamics is similar to the dynamics of a time-independent
multidimensional system. The QPKR can be simply realized experimentally
by exposing laser-cooled atoms (Cesium in the present work) to a delta-pulsed (kicked) laser standing
wave of wavenumber $k_L$ and time period $T_{1}$. The {\em amplitude} of the kicks
is quasiperiodically time-modulated with a frequency $\omega_{2}$.
The dynamics is effectively one-dimensional along the axis of the
laser beam, as transverse directions are uncoupled. The corresponding
Hamiltonian is: 
\begin{equation}
H=\frac{p^{2}}{2}+K\cos x\left[1+\varepsilon\cos\left(\omega_{2}t\right)\right]\sum_{n=0}^{N-1}\delta(t-n)\;,\label{eq:H}
\end{equation}
where $x$ is the particle position, $p$ its momentum, $K$ the kick
intensity and $\varepsilon$ the amplitude of the modulation. We have
chosen conveniently scaled variables such that
distances along the $x$ axis are measured in units of $(2k_L)^{-1}$,
the particle's mass
is unity and time is measured in units of the pulse period $T_{1}$. In
the quantum case, a crucial parameter is $\kbar\equiv4\hbar k_{L}^{2}T_{1}/M$,
the reduced Planck constant.

For $\varepsilon=0,$ one obtains the periodic kicked rotor, which
can be mapped onto a one-dimensional Anderson-like model~\cite{Fishman:LocDynAnderson:PRA84},
and displays ``dynamical'' localization \cite{Casati:LocDynFirst:LNP79,Altland:DiagrammaticAndersonLocQKR:PRL93},
that is, Anderson localization in momentum space instead of configuration
space. For non-zero $\varepsilon,$ 
the temporal dynamics of the QPKR is exactly that of a two-dimensional
periodic kicked system~\citep{Shepelyansky:Bicolor:PD87,Casati:IncommFreqsQKR:PRL89,Lemarie:AndersonLong:PRA09},
which itself can be mapped -- provided $2\pi/T_{1}$, $\omega_{2}$
and $\kbar$ are incommensurable numbers -- onto a two-dimensional Anderson-like 
anisotropic-hopping model, where anisotropy is controlled by $\varepsilon$ 
and the ratio of hopping to diagonal disorder is controlled
by $K/\kbar$~\footnote{See Supplemental Material at [URL will be inserted by publisher] for details 
on the mapping of the  $n$-frequency kicked rotor onto a $n$-dimensional Anderson model.}.

The existence of a mapping of the kicked rotor onto an Anderson-like
model has been used to experimentally observe 1D Anderson localization
with atomic matter waves as early as 1995~\cite{Moore:AtomOpticsRealizationQKR:PRL95}.
The two-frequency modulation of the QPKR -- which can be mapped on
a 3D Anderson model~\citep{Casati:IncommFreqsQKR:PRL89,Lemarie:AndersonLong:PRA09}
-- was used to experimentally observe 3D Anderson localization and
the metal-insulator Anderson transition~\cite{Chabe:Anderson:PRL08},
measure accurately the critical exponent and demonstrate its universality~\cite{Lopez:ExperimentalTestOfUniversality:PRL12}.

The experimental study of the 2D case is more challenging
than the 3D one, because the observation of the exponential behavior of
the localization length $p_{\mathrm{loc}}$ requires 
$p_{\mathrm{loc}}$ to vary over about one order of magnitude. The localization time increasing
with $p_{\mathrm{loc}}$, this also requires the ability to preserve
coherence over several hundreds kicks. This needed major evolutions of our experimental 
setup~\footnote{See Supplemental Material at [URL will be inserted by publisher] for details on the new standing wave setup.}.

Experimentally, an atomic sample consisting of few million atoms is prepared
in a thermal state (3.2 $\mu$K) whose momentum distribution is much
narrower than the expected localization length. The atomic cloud is
then ``kicked'' by a far-detuned ($\Delta\approx13$~GHz) pulsed
standing wave (SW). Pulse duration is typically $\tau=300$ ns, while
the typical pulse period $T_{1}=27.778$ $\mu$s corresponds to an
effective Planck constant $\kbar=2.89$. According to Eq.~(\ref{eq:H}),
an adjustable amplitude modulation with $\omega_{2}/2\pi=\sqrt{5}$
is superimposed to the kick sequence. In our previous experiments,
to minimize coupling with gravity, the SW was horizontal.
However, for 1000 kicks the atoms fall
down by 3.8 mm, compared to the 1.5 mm SW waist, limiting the maximum
number of kicks to 200. In order to overcome this
limit we used in the present experiment a {\em vertical} SW, and the atoms
fall freely between kicks. The new SW is formed by two beams that can be independently
controlled, both in amplitude and phase, through a radio-frequency
driving two acousto-optic modulators.
This allows us to accurately cancel gravity effects, by imposing a
linear frequency chirp to one arm of the SW with respect to the other, so that
the SW itself ``falls'' with acceleration $g$. A kicked rotor is
thus realized in the free-falling reference frame. 
These technical
improvements are discussed in more detail in the Supplemental Material~\cite{Note3}.
At the end of the sequence, the velocity
distribution is measured by a standard time-of-flight technique.

% Figure 1
\begin{figure}
\begin{centering}
\includegraphics[width=1\columnwidth]{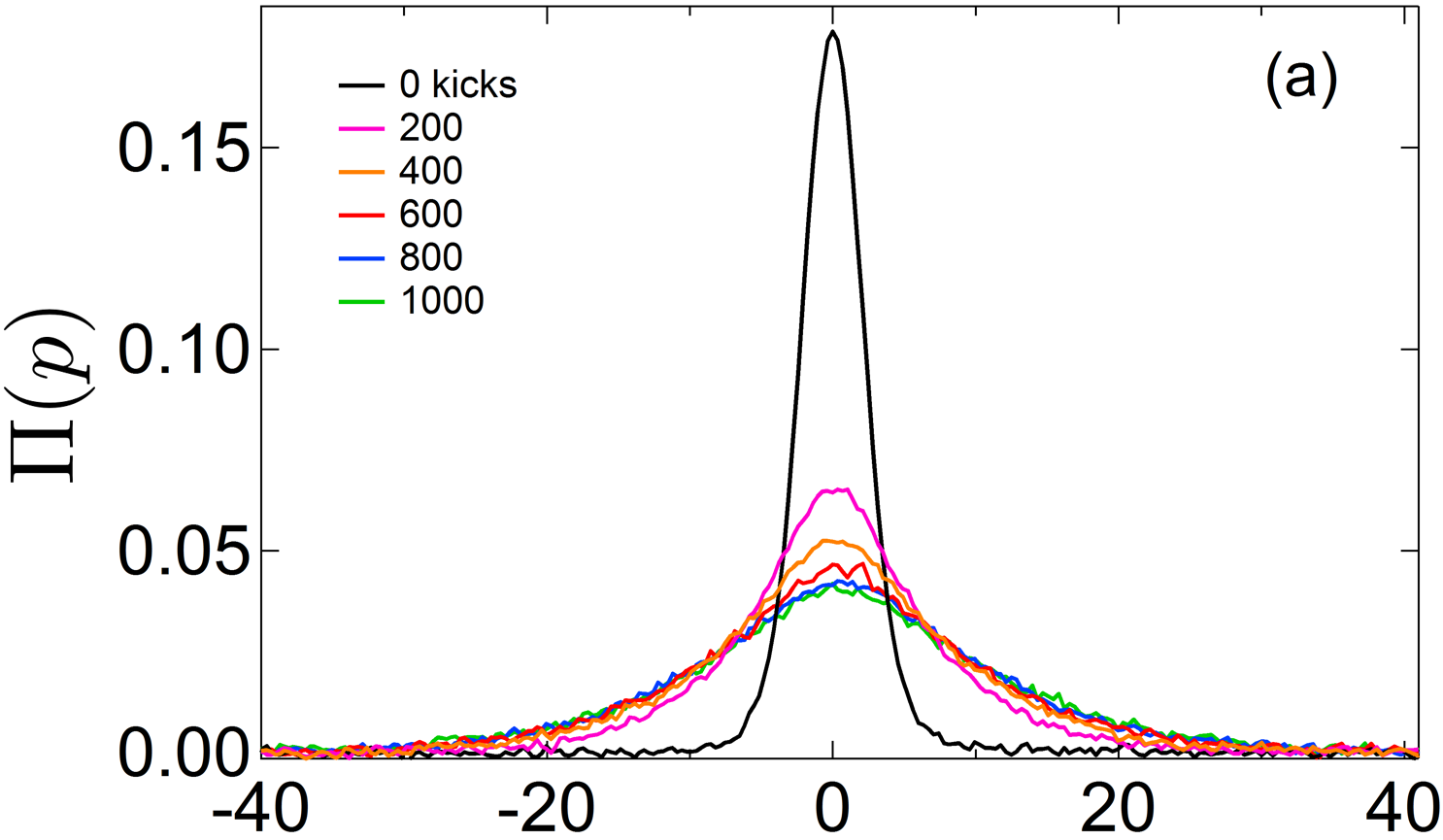} \\
\vspace{0.5cm}
\includegraphics[width=1\columnwidth]{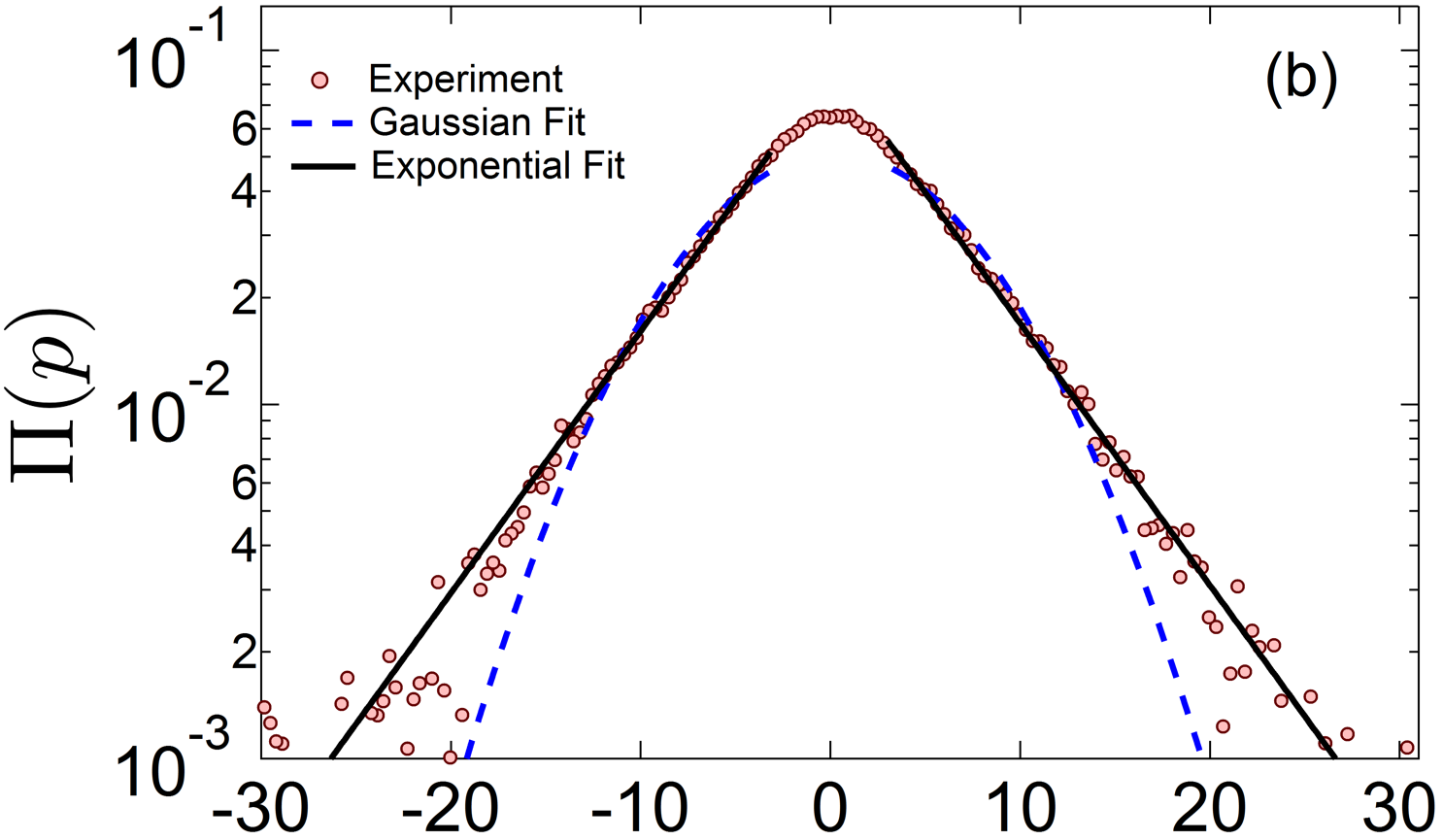} \\
\vspace{0.5cm}
\includegraphics[width=1\columnwidth]{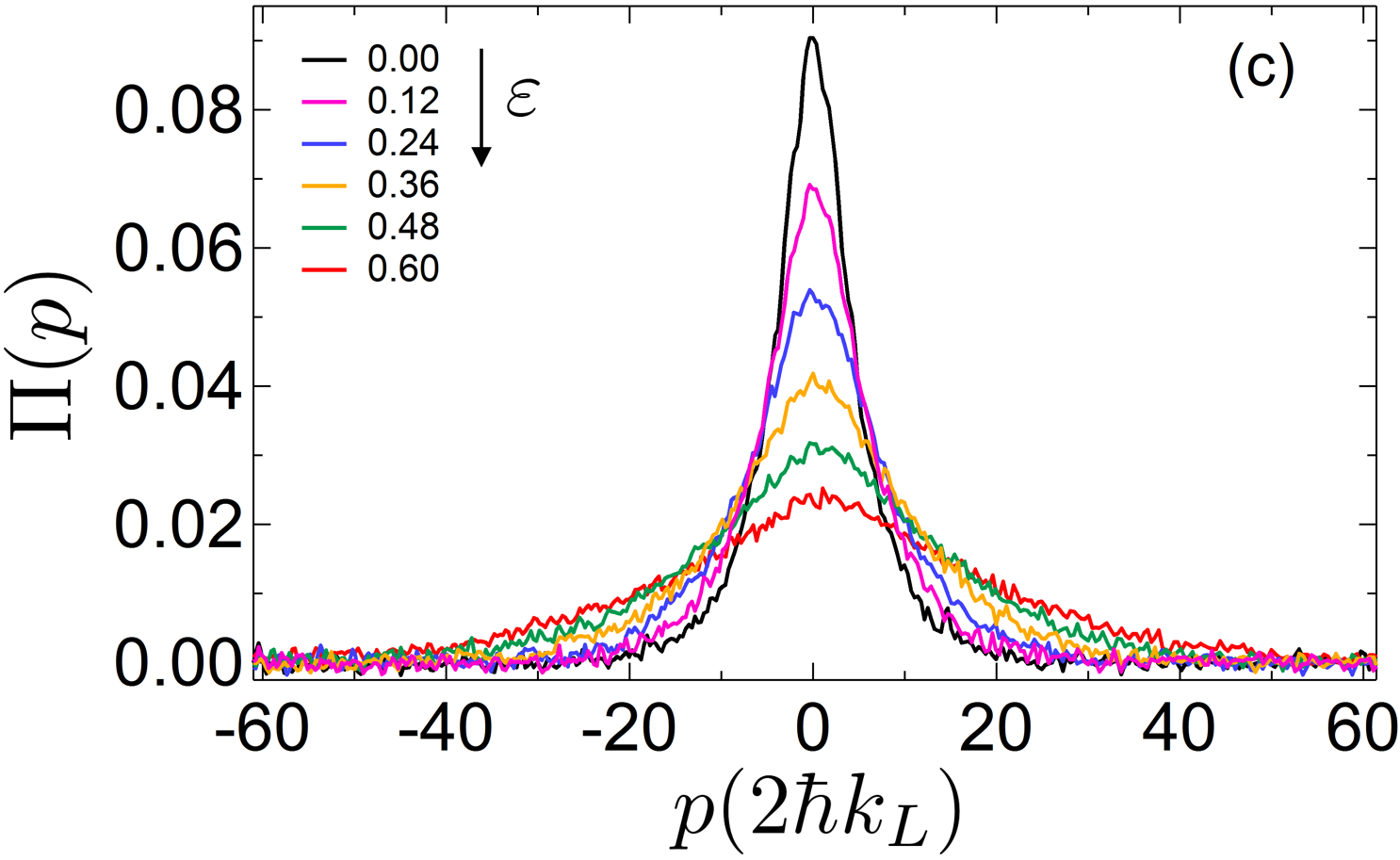} \\
\par\end{centering}
\protect\caption{\label{Fig:Momentum_distributions} (Color online) 
Experimentally recorded momentum distributions for the kicked rotor
exposed to a quasi-periodic driving, Eq.~(\ref{eq:H}).
(a) $K=5.34, \kbar=2.89,\varepsilon=0.36,$  0 to 1000 kicks (step 200). The momentum distribution
diffusively broadens at short times and freezes at longer times, proving the existence
of 2D Anderson localization. Time increases from top to bottom curves. (b) Momentum distribution 
at 200 kicks in log scale, showing the exponential shape characteristic of localization.
The circles are experimental points, the blue dashed line is a Gaussian fit
and the black solid line an exponential fit for $\left|p\right|>3\left(2\hbar k_{L}\right)$. 
(c) Localized momentum distributions after 1000 kicks, as a function of the anisotropy parameter $\varepsilon,$
for $K=5.34$, $\kbar=2.89$ as in (a) and (b).
The modulation amplitude
$\varepsilon$ increases from top to bottom curves. The rapid increase of the localization length
shows the evolution from the 1D localization at $\varepsilon=0$ to the truly 2D Anderson localization. 
Note the different horizontal scales in the various plots.}
\end{figure}

Figure~\ref{Fig:Momentum_distributions}(a) shows experimental momentum distributions $\Pi(p)$
recorded after 0 to 1000 kicks for  $K=5.34, \kbar=2.89, \varepsilon=0.36.$
If the dynamics were classical, the momentum distribution would keep
its initial Gaussian shape and the average kinetic energy increase
linearly with time, $E_{\mathrm{kin}}=E_{\mathrm{kin}}(t=0)+Dt$, where $D$ is the classical
diffusion constant in momentum space. 
In contrast, the experimental result displays a distribution which diffusively broadens at short times, but tends
to freeze, i.e. to \emph{localize} at long times. This clear-cut proof of localization is
confirmed by the shape of the momentum distribution, shown in Fig.~\ref{Fig:Momentum_distributions}(b) after
200 kicks. It very clearly displays an exponential shape~\footnote{The center of the momentum distribution is convoluted with the initial distribution, that is why it is excluded from the fits.}
 (a straight line in the logarithmic plot) $\exp\left(-|p|/p_{\text{loc}}\right)/2p_{\text{loc}}$
characteristic of localization with a localization length $p_{\text{loc}}$~\footnote{At longer times,
of the order of 1000 kicks, small deviations from the exponential profile are visible. They are
due to residual decoherence processes. We estimate the coherence time to be of the order of 400 kicks,
a significant improvement with respect to our previous setup.}. Figure~\ref{Fig:Momentum_distributions}(c) shows
the momentum distributions after 1000 kicks for  $K=5.34, \kbar=2.89$ and increasing
values of $\varepsilon$. It demonstrates that the localization length varies very rapidly with
$\varepsilon,$ indicating the evolution from a 1D localization at $\varepsilon=0$ to a truly 2D localization
with a much longer localization length at $\varepsilon=0.6.$ In order to prevent trivial
localization on KAM tori~\cite{Chirikov:ChaosClassKR:PhysRep79},
we always used $K>4,$ ensuring that the classical dynamics is ergodic.

Instead of measuring the full momentum distribution, it is sufficient
to measure the population $\Pi_{0}(t)$ of the zero velocity class as 
\begin{equation}
E_{\mathrm{kin}}\propto\frac{1}{4\Pi_{0}^{2}(t)}\label{eq:Ekin}
\end{equation}
is proportional to $\langle p^{2}\rangle(t)$ (as the total number
of atoms is constant) \footnote{We have also estimated directly the average kinetic energy from the
full momentum distribution and find similar results. The error bars
are however larger because of the difficulty to measure accurately
the tails of the distribution which contribute significantly.}.

%Figure 2
\begin{figure}
\begin{centering}
\includegraphics[width=1\columnwidth]{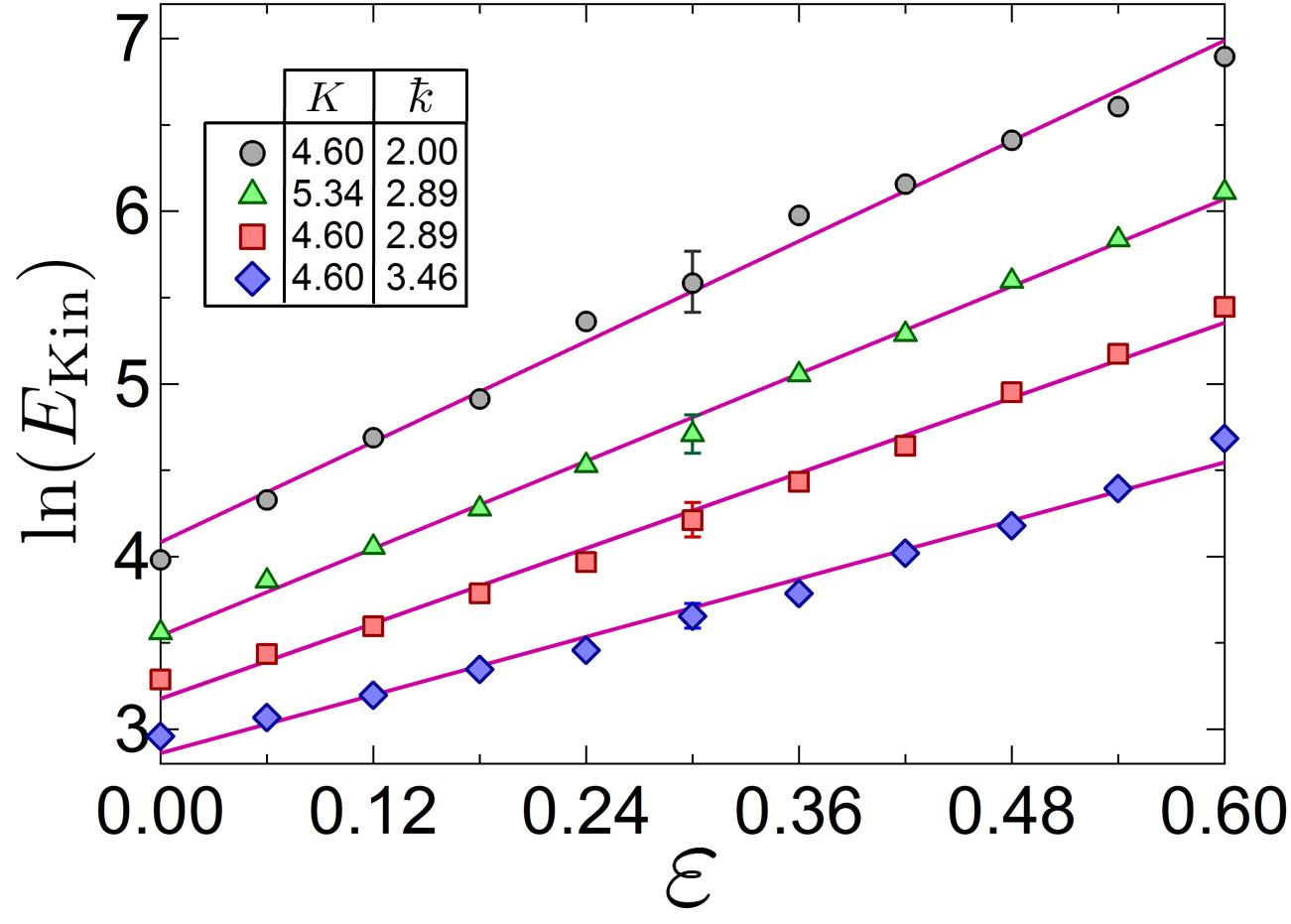} 
\par\end{centering}

\protect\caption{(Color online) Kinetic energy $E_{\mathrm{kin}}$ of the 
quasiperiodic kicked rotor vs. the modulation amplitude $\varepsilon,$ for 
various values of the kicking strength $K$ and effective
Planck constant $\kbar$.
The error bars indicate the typical experimental uncertainty. The
four curves are straight lines in this logarithmic scale, with a
slope that decreases with $\kbar$ and increases with $K.$ 
\label{Fig:Energy_vs_kbar} }
\end{figure}

Figure~\ref{Fig:Energy_vs_kbar} displays $E_{\mathrm{kin}}$ (at 1000
kicks) vs. $\varepsilon$ for various values of $K$ and $\kbar,$ 
showing that the exponential
dependence in $\varepsilon$ is a general feature, with a rate that
decreases with $\kbar$ and increases with $K$.
Note the overall $E_{\mathrm{kin}}$ dynamics of a factor of 60 (corresponding to a 8 fold increase in the localization
length), a key feature of the present experiment. 

The scaling theory of localization~\cite{Abrahams:Scaling:PRL79} predicts that dimension $d=2$ is
the lower critical dimension for the Anderson transition. For a time-reversal invariant spinless system 
(thus belonging to the \emph{orthogonal} universality class),
all states are localized with an exponentially large localization length.
For a usual 2D time-independent system, the relevant parameter is the dimensionless conductance at short scale,
equal to the product $k\ell$ of the wave vector by the mean free path, so that the logarithm of the localization
length is proportional to $k\ell$~\cite{Note1}.

The scaling theory cannot be directly transposed to the case of the kicked rotor for two reasons:
i) There is no wavevector playing the role of $k.$ Instead, one must consider the diffusion constant (in momentum
space), which is, for a periodic kicked rotor, approximately equal to $K^2/4$. 
ii) The diffusion process for the 2D quasiperiodic kicked rotor is not isotropic. As shown in
\cite{Lemarie:KR3DClassic:JMO10} and discussed in the Supplemental Material~\cite{Note1},
the quasiperiodic kicked rotor
can be mapped on a 2D Anderson-like model, whose dynamics at short time is indeed diffusive, but anisotropic.
Along the ``physical'' direction (which coincides with
the atom momentum component along the standing wave), the diffusion constant
is -- for small $\varepsilon$ -- almost equal to the one of the periodic kicked rotor, $D_{11} \approx K^2/4$;
along the other (virtual) direction, the diffusion constant is $D_{22} \approx K^2\varepsilon^2/8$,
so that it vanishes in the limit $\varepsilon \to 0,$ where one must recover the usual 1D periodic kicked rotor.

Altogether, the relevant parameter is the geometric average of the diffusion constant along the two directions
$\sqrt{D_{11}D_{22}}\propto \varepsilon K^2$. The scaling theory predicts that the logarithm
of the localization length should be proportional to $\sqrt{D_{11}D_{22}}/\kbar^2$. A similar prediction was made
in~\cite{Shepelyansky:Bicolor:PD87} using a slightly different method.

%Figure 3
\begin{figure}
\begin{centering}
\includegraphics[width=1\columnwidth]{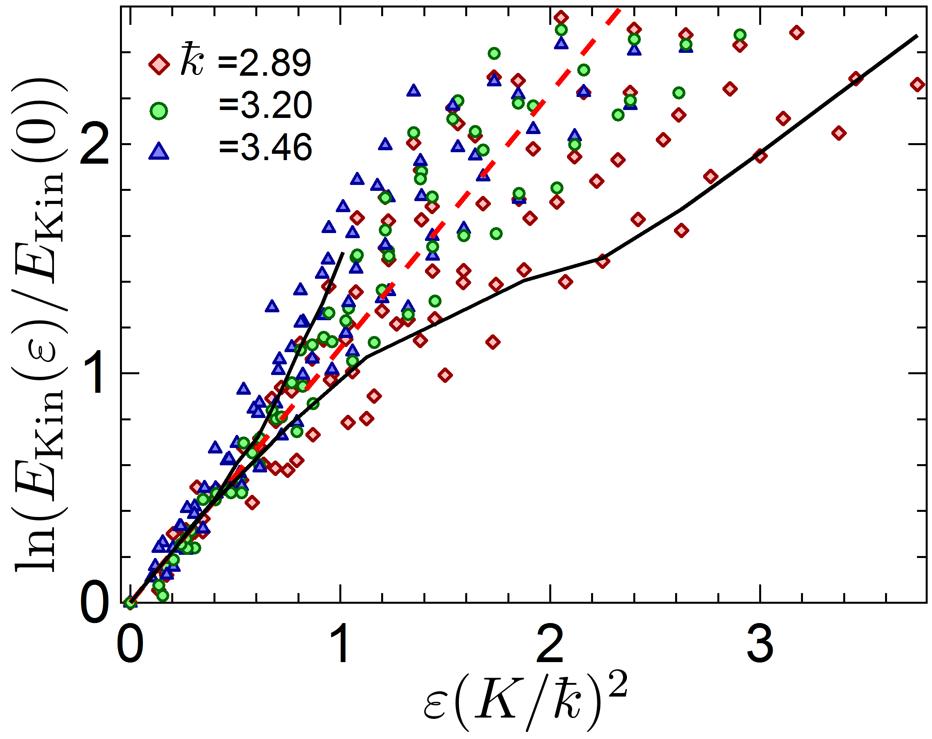} 
\end{centering}
\protect\caption{\label{Fig:Nuage_de_points} (Color online) Increase in 
the kinetic energy at $t=1000$ ($\propto p_\mathrm{loc}^2$) of
the quasiperiodic kicked rotor with respect to the purely one-dimensional
situation $\varepsilon=0$ vs.~the scaling parameter $\varepsilon(K/\kbar)^2.$
The cloud of experimental points -- collected at various values of $K,$
$\varepsilon$ and $\kbar$ -- is distributed around an average
linear dependence in this semi-logarithmic plot, which shows the exponential dependence
of the localization length, characteristic of 2D Anderson localization. 
The red dashed line is the prediction of Eq.~(\protect \ref{eq:prediction_xiloc}). 
The spread is due
in part to experimental imperfections (at large $\varepsilon(K/\kbar)^2,$ the localization time
is not much shorter than the duration of the experiment) and in part to fundamental reasons:
The linear dependence on $\varepsilon K^2/\kbar^2$ in the argument of the exponential, 
Eq.~(\protect \ref{eq:prediction_xiloc}),
is valid only at small $\varepsilon$, and the formula assumes that the classical diffusion 
constant is proportional to $K^2,$ while the actual diffusion constant has oscillatory corrections.
The black curves are numerical simulations corresponding to the two 
``extreme'' values of $K/\kbar=1.3, \kbar =3.46$ (higher curve)
and $K/\kbar=2.5, \kbar=2.89$ (lower curve); they display the same spreading phenomenon.}
\end{figure}

The self-consistent theory of localization is an attempt towards more quantitative predictions,
based on the same ideas as the scaling theory.
It has been successfully used to predict properties of the Anderson transition~\cite{Kroha:SelfConsistentTheoryAnderson:PRB90,Zambetaki:LocalizationAnisotropic:PRL96},
and was transposed to the periodic kicked rotor in~\cite{Altland:DiagrammaticAndersonLocQKR:PRL93,Tian:EhrenfestTimeDynamicalLoc:PRB05}
and to the quasiperiodic kicked rotor with two additional driving
frequencies in~\cite{Lopez:PhaseDiagramAndersonQKR:NJP13}.
It consists in computing perturbatively the weak localization correction
to the (anisotropic) diffusion constant and to extrapolate to the
strong localization regime. It however depends on the cut-offs used.
For our quasiperiodic kicked rotor~\cite{Note1} it confirms the prediction of the scaling theory,
namely:
\begin{equation}
p_{\text{loc}}=\frac{K^{2}}{4\kbar}\exp\left(\frac{\alpha\varepsilon K^{2}}{\kbar^{2}}\right)\label{eq:prediction_xiloc}
\end{equation}
where $\alpha$ is a number of the order unity, which may however depend smoothly
on the parameters. In the limit $\varepsilon\to 0,$ it is 
$\alpha = \pi/\sqrt{32}.$

In Fig.~\ref{Fig:Nuage_de_points}, we display the results of 275
measurements, corresponding to 12 values of the ratio $K/\kbar\in[1.3,2.5]$,
with $K\in[4.33,7.26]$ and $\kbar=\left\{ 2.89,3.2,3.46\right\} $,
and to $\varepsilon$ values from 0 to 0.6 (step 0.06). Dividing $E_{\mathrm{kin}}(\varepsilon)$
by $E_{\mathrm{kin}}(\varepsilon=0)$ makes it possible to probe the
exponential term in Eq.~(\ref{eq:prediction_xiloc}). The exponential dependence (straight line in logarithmic scale) 
is visible for $\varepsilon\lesssim 1$,
materialized by the red dashed line, corresponding to the prediction $\alpha=\pi/\sqrt{32}$ of the
self-consistent theory. Despite the
spreading of the experimental results around the average trend, the overall agreement is rather good.
This proves the exponential dependence of the localization length in 2D,
and thus that $d=2$ is the lower critical dimension for the metal-insulator Anderson
transition. 
Some deviations are nevertheless visible. They arise from different phenomena: First, for large
$\varepsilon$, the localization time can be only slightly
shorter than the duration of the experiment (1000 kicks), meaning
that the measured momentum distribution is not the asymptotic one
for infinite time and underestimates the eventual saturation of $E_{\mathrm{kin}}$
at long time. This explains why the experimental points at large $\varepsilon$
tend to lie below the theoretical prediction. This is confirmed by numerical calculations in the experimental conditions for the 
largest value of $K/\kbar = 2.5$ (longest localization time), see the solid lower curve in Fig.~\ref{Fig:Nuage_de_points}.
A second, more fundamental, phenomenon is that Eq.~(\ref{eq:prediction_xiloc}) assumes
that the classical diffusion constant is simply $K^{2}/4,$ which
is valid only for $K\gg1$, whereas oscillatory corrections at moderate
$K$ are known to exist for the 1D kicked rotor \cite{Shepelyansky:Kq:PD83}
and to persist even for the 3D QPKR~\cite{Lemarie:KR3DClassic:JMO10}. This dependence is thus not eliminated
by the normalization to $E_\mathrm{kin}(\varepsilon=0)$. This explains a significant part of the spreading of the data.
Finally, Eq.~(\ref{eq:prediction_xiloc}) is expected to be valid in the $\varepsilon\to 0$ 
limit, see Supplemental Material~\cite{Note1}. At larger 
$\varepsilon$ values, higher order terms must come into play and are responsible for significant deviations. This is visible in Fig.~\ref{Fig:Nuage_de_points},
where both experimental (points) and numerical (solid lines) data are well predicted at small 
$\varepsilon K^{2}/\kbar^{2},$ but are more widely spread
as $\varepsilon K^{2}/\kbar^{2}$increases. A thorough analysis of all these deviations is beyond the scope of this Letter.

To summarize, we presented the first experimental evidence of
two-dimensional Anderson localization with atomic matter waves. We
studied the variation of the localization length with the system
parameters and showed that it displays an exponential dependence characteristic
of time-reversal spinless systems. To the best of our knowledge, such
an experimental evidence has not been observed previously. It demonstrates experimentally that $d=2$
is the lower critical dimension of the Anderson transition. The observed localization
length varies as predicted by the scaling and the self-consistent theories of localization.

\begin{acknowledgments}
The authors acknowledge M. Lopez for his help in the early stages
of this experiment. DD thanks N. Cherroret, S. Ghosh and C. Tian for discussions
on the self-consistent theory of localization. This work was supported
by Agence Nationale de la Recherche (Grants LAKRIDI No. ANR-11-BS04-0003
and K-BEC No. ANR-13-BS04-0001-01), the Labex CEMPI (Grant No. ANR-11-LABX-0007-01),
and ``Fonds Europ{\'e}en de D{\'e}veloppement Economique R{\'e}gional''
through the ``Programme Investissements d'Avenir''. This work was
granted access to the HPC resources of TGCC under the allocation 2015-057083
made by GENCI (``Grand Equipement National de Calcul Intensif'') and to
the HPC resources of The Institute for Scientific Computing and Simulation
financed by Region Ile de France and the project Equip@Meso (Grant No.
ANR-10-EQPX- 29-01). 
\end{acknowledgments}

\appendix*
\section{Supplemental Material}
The following Supplemental Material discusses: 

\begin{enumerate}[I.]
\item The main improvements in our experimental setup allowing the 2D Anderson localization to be observed.
\item The mapping of the quasiperiodic kicked rotor onto a 2D Anderson model, indicating the essentials of this mapping and its peculiarities relevant for the present work.
\item The classical diffusion in the 2-frequency quasiperiodic kicked rotor.
\item A schematic derivation of the localization length, indicating the essential steps of the calculation in the case of the quasiperiodic kicked rotor with 2 incommensurate frequencies and the origin of the
exponential dependence on the anisotropy.
\end{enumerate}

\subsection{Setup improvements}

\begin{figure}
\begin{centering}
\includegraphics[width=\columnwidth]{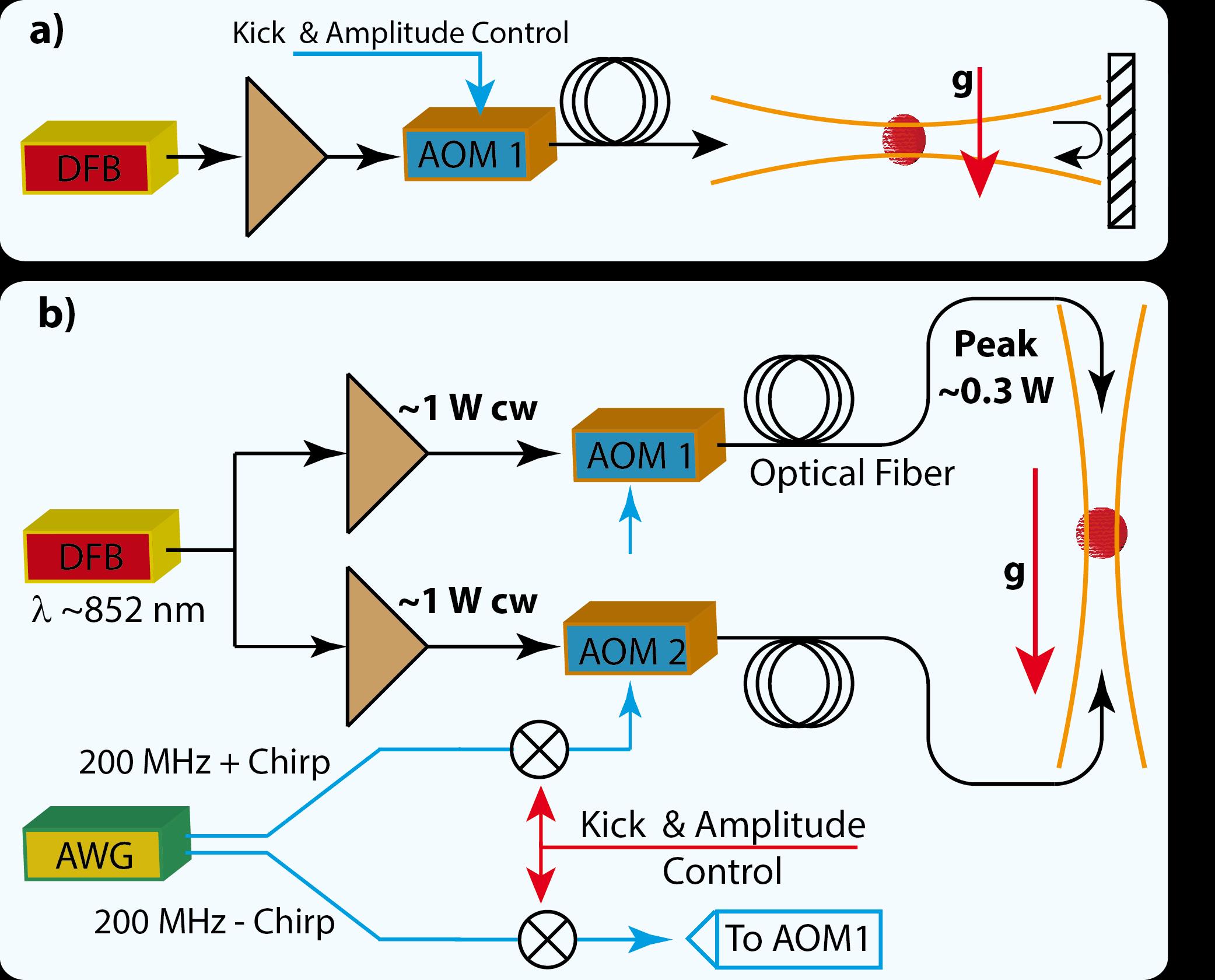} 
\par\end{centering}

\protect\caption{ \label{Fig:NewSW}Evolution of the experimental setup.
a) Horizontal standing wave setup used in former experiments. The kicks are controlled
by a single acousto-optic modulator (AOM), and the standing wave is built by
retro-reflecting the incoming laser beam. b) Schematic view of the
new vertical setup used in the present work. 
A Distributed Feedback
Laser Diode (DFB) seeds two optical amplifiers. 
The kick-sequence temporal modulation is produced by an arbitrary
wave generator (AWG). In addition to the pulsed sequence, 
a linear chirp is added so that
the standing wave ``falls'' with acceleration $g$,
simultaneously with the atomic cloud.}
\end{figure}

In our previous experiments, in order to minimize coupling with gravity,
the standing wave (SW) was horizontal, see Fig.~\ref{Fig:NewSW}a. This
geometry has several drawbacks. For 1000 kicks, the atoms
would fall of 3.8 mm, to be compared to the 1.5 mm of the SW waist.
Moreover, the SW was built by retro-reflecting the incoming beam,
which leads to a $\sim$1 m optical path difference between the two
beams overlapping on the atomic cloud region, a source of phase noise
detrimental to dynamical localization. These restrictions limited
the kick number to less than 200 in our previous experiments. In order
to overcome this limit, we built a new SW system with several improvements,
as shown in Fig.~\ref{Fig:NewSW}b. The SW is now vertical, and,
between kicks, the atoms fall freely. The intensity inhomogeneity
in the transverse direction was also reduced.
The SW is now formed by two independent
beams, which has many advantages, as each arm can be independently
controlled, both in amplitude and phase through the radio-frequency
wave that drives the acousto-optic modulators. This allows us
to accurately cancel gravity effects, by imposing a linear frequency
chirp to one of the arms with respect to the other, so that the SW
itself ``falls'' with acceleration $g$. A kicked rotor is thus
realized in the free-falling reference frame. Finally, the SW phase
noise induced by the laser linewidth is minimized by accurately balancing
the optical paths of the arms to better than 1 cm. This is performed
by directly minimizing the kinetic energy dispersion at 1000 kicks
with a 1D kicked rotor.

%Figure 1

\subsection{Mapping of the quasiperiodic kicked rotor onto a 2D Anderson model}
The mapping of the quasiperiodic kicked rotor (QPKR) onto a 2D Anderson model has been described in~\citep{Shepelyansky:Bicolor:PD87,Casati:IncommFreqsQKR:PRL89,Lemarie:AndersonLong:PRA09}. Because it is useful to derive an approximate
expression for the localization length, we summarize here the basics of the calculation.

The starting point is the following Hamiltonian:
 \begin{equation}
\mathcal{H}=\frac{p_{1}^{2}}{2}+\omega_{2}p_{2}+K\cos x_{1}\left[1+\varepsilon\cos x_{2}\right]\sum_{n}\delta(t-n)
 \label{eq:KR2DquasiperH}
 \end{equation}
which describes a {\em periodically} kicked, two-dimensional ``rotor''; we  use here quotations marks around ``rotor'' because
of the unusual form of the kinetic energy along the direction 2, linear instead of quadratic.
It is a matter of algebra~\cite{Lemarie:AndersonLong:PRA09}
to show that the temporal evolution of an initial state initially localized 
at $x_2=\phi_2,$ but with arbitrary wave-function along $x_1$:
\begin{equation}
\label{eq:Psi3}
 \Psi({x}_{1},{x}_{2},t=0)\equiv\Xi({x}_{1},t=0)\delta({x}_{2}-\phi_2)
\end{equation}
leads to the state:
\begin{equation}
 \Psi({x}_{1},{x}_{2},t) = \Xi({x}_{1},t)\delta({x}_{2}-\phi_{2}-\omega_{2}t),
\label{eq:2d}
\end{equation}
where the wavefunction $\Xi({x}_{1},t)$ \emph{exactly} obeys the time-dependent Schr\"odinger
equation of the quasiperiodic kicked rotor:
\begin{equation}
H=\frac{p_1^2}{2}+K\cos x_1\left[1+\varepsilon\cos\left(\omega_{2}t+\phi_2\right)\right]\sum_{n}\delta(t-n)\;.
\label{eq:H1}
\end{equation}
Thus, the evolution of the quasiperiodic kicked rotor can be thought of as the evolution 
of a 2D periodic ``rotor'', with a peculiar initial state perfectly localized in the $x_2$ direction and thus
completely delocalized in the conjugate $p_2$ direction.

The Hamiltonian Eq.~(\ref{eq:KR2DquasiperH}) being time-periodic, one can use the Floquet theorem and look at its
Floquet eigenstates. Let $|\phi\rangle$ denote an eigenstate of the evolution operator over one period
with eigenvalue $\exp(-iE/\kbar),$ and consider the state $|\chi\rangle=(1+iW)^{-1}|\phi\rangle$
where $W(x_1,x_2)\equiv\tan\left[K\cos{x}_{1}(1+\varepsilon\cos{x}_{2})\right/2\kbar].$ 
The Hamiltonian being periodic in $x_1$ and $x_2$ (we take periodic boundary conditions),  $|\chi\rangle$
can be expanded on the basis of plane waves $|m_1,m_2\rangle$ -- eigenstates of the momentum operators
in both directions with eigenvalues $m_1\kbar$ and $m_2\kbar$: $|\chi\rangle = \sum_{m_1,m_2}{\chi_{m_1,m_2}|m_1,m_2\rangle}.$ 
It is again a matter of algebra~\cite{Casati:IncommFreqsQKR:PRL89,Fishman:LocDynAnderson:PRA84} to show that the coefficients
$\chi_{m_1,m_2}$ obey the following equation:
\begin{equation}
\epsilon_{m_1,m_2}\chi_{m_1,m_2}+\sum_{r_1,r_2\neq0}W_{r_1,r_2}\chi_{m_1+r_1,m_2+r_2}=0,
\label{eq:AndersonmodelKRquasiper}
\end{equation}
where $\epsilon_{m_1,m_2}$ is given by:
\begin{equation}
\epsilon_{m_1,m_2}=\tan\left\lbrace \frac{1}{2}\left[\left(\kbar\frac{{m_{1}}^{2}}{2}+\omega_{2}m_{2}\right)-\frac{E}{\kbar}\right]\right\rbrace 
\label{eq:pseudo-random-disorder-quasiperKR}
\end{equation}
and the $W_{r_1,r_2}$ are the two-fold Fourier components of the doubly-periodic function $W.$

Equation~(\ref{eq:AndersonmodelKRquasiper}) can be interpreted as the eigenvalue equation for a Anderson-like model
on a 2D lattice with sites labeled $(m_1,m_2),$ with hopping described by $W_{r_1,r_2}$ and on-site energies $\epsilon_{m_1,m_2}.$ 
There are however four differences with respect to a usual Anderson model:
\begin{itemize}
 \item The hopping matrix elements $W_{r_1,r_2}$ are not limited to nearest neighbors. They do however
decrease fast enough at large $(r_1,r_2)$ so that this difference does not change the qualitative behavior
(i.e. localization).
\item The on-site energies $\epsilon_{m_1,m_2}$ are not random, but rather a deterministic pseudo-random sequence.  
Provided $\pi,\kbar,\omega_2$ are incommensurate, the sequence has no periodicity and localization properties similar
to those of a truly random model are expected (and observed numerically).
\item The hopping is anisotropic, governed by the $K/\kbar$ coefficient along direction 1 and by the  $\varepsilon K/\kbar$ coefficient along direction 2. For $\varepsilon=0,$ the 2D system appears as a series of chains
along direction 1, which are uncoupled along direction 2, and a 1D Anderson model is recovered.
\item Floquet quasi-eigenstates with different quasi-energies $E$ are associated with the same energy 0 in Eq.~(\ref{eq:AndersonmodelKRquasiper}), but with different realizations of the disorder, Eq.~(\ref{eq:pseudo-random-disorder-quasiperKR}), all having the same statistical properties. This is why, for given values of parameters $\kbar,K,\varepsilon,$ all quasi-eigenstates have the same localization length \cite{Note1}.
\end{itemize}

In order to understand localization properties of the quasiperiodic kicked rotor, it is thus sufficient to study
transport and localization on the equivalent anisotropic Anderson-like 2D model Eq.~(\ref{eq:KR2DquasiperH}).
 
\subsection{Classical diffusion}

It is well known that the classical dynamics of the periodic kicked rotor is described by the Standard Map~\citep{Chirikov:ChaosClassKR:PhysRep79}.
A similar map can be constructed for the quasiperiodic kicked rotor. The classical evolution over one period
of the Hamiltonian Eq.~(\ref{eq:H1}) is given by the following map:
\begin{eqnarray}
p_{1_{n+1}} & = & p_{1_{n}}+K\sin x_{1_{n}}(1+\varepsilon\cos x_{2_{n}})\;,\nonumber \\
p_{2_{n+1}} & = & p_{2_{n}}+K\varepsilon\cos x_{1_{n}}\sin x_{2_{n}}\;,\nonumber \label{eq:KR3DApplicationStandard}\\
x_{1_{n+1}} & = & x_{1_{n}}+p_{1_{n+1}}\;,\nonumber \\
x_{2_{n+1}} & = & x_{2_{n}}+\omega_{2}\;,\nonumber
\end{eqnarray}
where $y_n=y(t=n+\epsilon)$, $y=x_i,p_i$ ($i=1,2$).
If $K$ is sufficiently large, the classical dynamics is almost fully chaotic~\cite{Lemarie:KR3DClassic:JMO10}. For the 1D problem
($\varepsilon=0$), this takes place for $K\gtrsim 6$. For the 2D problem (non vanishingly small $\varepsilon$), we found an almost fully chaotic dynamics for $K\gtrsim4$, and this is why we did not perform any experiment below this value.
In the chaotic regime, the kicks in momentum have a pseudo-random sign (depending on $x_1$ and $x_2$), making the
dynamics in momentum space appear as a pseudo-random walk, i.e. a chaotic diffusive process at long time.
Note however that the diffusion is anisotropic because kicks along $p_2$ are typically smaller than kicks along $p_1$ by a factor
$\varepsilon.$  
In the limit of large $K$ it is easy to evaluate the diffusion tensor, defined as
\begin{equation}
D_{ij} = \lim_{t\rightarrow +\infty} \frac{\langle p_i p_j \rangle}{t}
\end{equation}
by assuming that positions of consecutive kicks are
uniform uncorrelated variables (see~\cite{Lemarie:KR3DClassic:JMO10} for the essentially identical
calculation in 3D). It is diagonal in the (1,2) directions with:
\begin{eqnarray}
D_{11} & \approx & (K^{2}/4)(1+\varepsilon^{2}/2)\;,\nonumber\\
D_{22} & \approx & K^{2}\varepsilon^{2}/8\;,\nonumber\\
D_{i\neq j} & \approx & 0\;.
\label{diff:tensorD}
\end{eqnarray}
Numerical simulations of the classical dynamics~\cite{Lemarie:KR3DClassic:JMO10} fully confirm that the
classical dynamics is an anisotropic diffusion; however, at not too large $K,$ correlations between successive kicks
are responsible for oscillatory corrections to the diffusion constant of the order of few ten percents.

\subsection{Localization length}
The calculation of the localization length turns out to a bit tricky in 2D. It has been discussed in the literature
mainly in the context of electrons in disordered potentials~\cite{Vollhardt:SelfConsistentTheoryAnderson:92,Woelfle:SelfConsistentTheoryAndersonLoc:IJMPB10}. 
The calculation for the quasiperiodic rotor follows the same lines.
We sketch here only the general method, comparing the key steps with 
the case of electrons in disordered potentials.

The starting point is to take into account the weak localization effect due to closed loops in the system. In the weak
scattering regime, where this correction is small, it takes the following form for electrons (or massive particles)
in disordered potentials:
\begin{equation}
 \frac{{\cal D}}{D} = 1 - \frac{1}{\pi \rho} \int{\frac{1}{(2\pi)^2} \frac{\rmd \vecq}{D q^2}}
\label{Eq:WL_electrons}
\end{equation}
where $D$  is the classical (Boltzmann) diffusion constant and ${\cal D}$ the quantum modified one. 
The term after 1 in
the right-hand side is the weak localization correction. It depends on the density of states $\rho$ and involves a two-dimensional
integral over the vector $\vecq$ which is a momentum, conjugate of position. The integrand is nothing
but the classical diffusive kernel (at zero frequency, hence infinite time). Formally, the integral diverges both for small and large $q,$ so that appropriate cut-offs must be set. The natural short distance (large $q$) cut-off is the mean free path $\ell$ 
(at shorter scale, the dynamics is not diffusive). The short $q$ cut-off can be taken 
proportional to $1/L$, where $L$ is the size of the system.
One then obtains a size-dependent diffusion constant ${\cal D}(L).$ Taking for the density of states in 2D its disorder-free value $1/2\pi$
(we take the mass of the particle and $\hbar$ as unity), one obtains:
\begin{equation}
\frac{{\cal D}(L)}{D} = 1 - \frac{2}{\pi k \ell} \ln\left(\frac{L}{\ell}\right)
\label{Eq:DL_electrons}
\end{equation}
where $k$ is the wavevector of the particle. In these units, the classical diffusion constant is $D=k\ell/2.$
The logarithmic dependence in $L$ is a crucial ingredient, as we will see below. It is intimately related to
the low-$q$ divergence of the integral. 

Equation~(\ref{Eq:DL_electrons}) shows that the diffusion constant decreases with the system size. Of course,
this cannot be correct for arbitrary large size, as the correction is computed assuming it is small. This ceases
to be true when the right hand side term in Eq.~(\ref{Eq:DL_electrons}) vanishes. This gives an order of magnitude
of the size at which the diffusion constant vanishes, that is the localization length. One then gets 
\begin{equation}
\xi=\ell\exp\left(\frac{\pi k\ell}{2}\right).
\end{equation}
Of course, this is only a very approximate expression. The self-consistent theory of localization~\cite{Vollhardt:SelfConsistentTheoryAnderson:92,Woelfle:SelfConsistentTheoryAndersonLoc:IJMPB10} is an attempt to be a bit more quantitative. It predicts essentially the same exponential dependence for the localization length. 
 
For the periodic kicked rotor, the weak localization corrections have been computed in~\cite{Altland:DiagrammaticAndersonLocQKR:PRL93}. These results have been extended to the quasiperiodic rotor in~\cite{Lopez:PhaseDiagramAndersonQKR:NJP13,Tian:TheoryAndersonTransition:PRL11}. 
There are essentially two modifications:
\begin{itemize}
\item The term depending on the density of states (which is meaningless for a time-periodic system where
the density of states of the Floquet Hamiltonian is infinite) $1/\pi\rho$ must be replaced by $2\kbar^2.$
\item Because the classical dynamics is an anisotropic diffusion, the diffusive
kernel is now $(D_{11}q_1^2+D_{22}q_2^2)^{-1}$. When performing the integral over $\vecq,$ it is not entirely
clear how to choose the cut-offs. The simplest choice is to take the large $q_1$ and $q_2$ cut-offs scaling like
the (anisotropic) mean free paths, i.e. respectively proportional to $1/\sqrt{D_{11}}$ and $1/\sqrt{D_{22}}$. This is however arbitrary
and questionable. It is important to understand that the choice of different cut-offs will affect
the prefactors, but not the key point, namely the logarithmic dependence of the integral on the system size. 
\end{itemize}
With this simple choice of cut-offs, the weak localization correction reads:
\begin{equation}
 \frac{{\cal D}_{11}(L)}{D_{11}} = \frac{{\cal D}_{22}(L)}{D_{22}} = 1 - \frac{\kbar^2}{\pi\sqrt{D_{11}D_{22}}} \ln\left(\frac{L}{l}\right) 
\label{Eq:WL_KR}
\end{equation}
where $l$ is a yet unspecified short scale cut-off.

With the same reasoning as the one for electrons, one obtains the approximate expression for the localization length
(in momentum space for the kicked rotor):
\begin{equation}
p_{\mathrm{loc}} = l \exp\left(\frac{\pi \sqrt{D_{11}D_{22}}}{\kbar^2}\right). 
\label{Eq:xi_KR}
\end{equation}
In the $\varepsilon \to 0$ limit, one must recover the localization length of the periodic kicked rotor, which fixes 
$l=K^2/4\kbar$.
By inserting the values of the diffusion constants, Eqs.~(\ref{diff:tensorD}) into Eq.~(\ref{Eq:xi_KR}),
we finally obtain:
\begin{equation}
p_{\mathrm{loc}}=\frac{K^{2}}{4\kbar}\exp\left(\frac{\pi\varepsilon K^{2}}{\sqrt{32}\kbar^{2}}\right)\,.
\label{eq:prediction_xiloc1}
\end{equation} 

Because of the questionable assumptions on the cut-offs, it might well be that the coefficient $\pi/\sqrt{32}$ is not exact
and requires some correction. This is a difficult question left for further investigation. We just
emphasize that the exponential behavior of the localization length with the scaling parameter $\varepsilon K^{2}/\kbar^{2}$
comes from the prefactors in the integral over $\vecq$ and from the logarithmic singularity of the integral. It is thus
a very robust phenomenon, unlikely to be affected by the shortcomings of the self-consistent theory of localization.
Ultimately, the \emph{experimentally observed} exponential dependence of the localization length
\emph{proves} that the weak localization correction is logarithmic with the system size. 
This, in turn, implies that dimension $d=2$ is the lowest critical dimension of the Anderson transition.
Indeed, a modification of the dimension would introduce an extra $q^{d-2}$ factor in the integral, i.e.
would change the logarithm dependence in $L$ in Eq.~(\ref{Eq:WL_KR}) to an algebraic one, incompatible with
the exponential behavior of the localization length.

%\bibliographystyle{apsrev} Incompatible with the option `longbibliography'
%\bibliography{ArtDataBase_v2}

\end{document}